\documentclass[showpacs,preprintnumbers,amsmath,amsthm,amssymb,aps]{revtex4}
\usepackage[dvips]{graphicx}

\begin{document}

\title{Study Zitterbewegung effect in a Quasi One-dimensional Relativistic Quantum Plasma by DHW formalization}
\date{\today}%

\author{Safura Nematizadeh Juneghani$^{1}$}%
\affiliation{$^{1}$Department of Physics, Shahid Beheshti University, Tehran, Iran}%
\author{Babak Shokri$^{1,2}$}%
\email{b-shokri@sbu.ac.ir}
\affiliation{$^{1,2}$Laser and Plasma Research Institute, Shahid Beheshti University, Tehran, Iran}%
\begin{abstract}%
Using Dirac equation together with the Wigner distribution
function, the trembling motion, known as Zitterbewegung effect, of
moving electrons in quasi-one-dimensional relativistic quantum
plasma is theoretically investigated. The relativistic Wigner matrix
is used to calculate the mean values of the position and velocity
operators for a Dirac gas of electrons. It is found
that the oscillatory behavior of measurable quantities could be
associated with the Zitterbewegung effect which manifests itself
as a damping interference pattern stemming from mixing the
positive and negative dispersion modes of Dirac particles.
\end{abstract}
\maketitle%
\section{INTRODUCTION}%
\label{sec1}%
\indent%
A relativistic quantum plasma, a plasma under particular
conditions, could be achievable when the Fermi energy and the de
Broglie wavelength (or even the Compton length) of its ingredients
become comparable to or higher than their rest energies and the
average inter-particle separations, respectively. In addition to Bohm and
exchange-correlation potential, the relativistic speeds of Fermions
play crucial roles in the system
\cite{R1,R2,R3,R4,R5,R6,R7,R8,R9,R10,R11}. Since the early work by
Chandrasekhar\cite{R12,R13}, this kind of plasma has been
established in astrophysical compact objects like white dwarf and
neutron stars, and in the matter of the laboratory plasma as well
\cite{R14}. In quantum plasmas, the degeneracy parameter is measured
via the number density of the Fermi particles. Also, in degenerate plasmas
when carriers are travelling with relativistic velocities, the
degeneracy parameter gets modified \cite{R8,R9}.\\
\indent%
More significantly, motivated by the recent results obtained with
trapped ions \cite{R15}, it is now becoming possible to simulate
Dirac-like Hamiltonian so as to examine relativistic quantum effects
such as Zitterbewegung (ZB), a quivering and an astonishing motion
of a free relativistic particle, which was first proposed by
Schrodinger \cite{R16,R17,R18,R19,R20,R21}. In quantum simulation, a
quantum system could be imitated by a controllable laboratory system
that accounts for the same mathematical form \cite{R15}. In this way,
a multitude of studies have been performed to examine the elusive
trembling motion of carriers in systems with gap in energy
spectra like exciton-polaritons in two-dimensional
microcavities \cite{R22}, phononic crystals \cite{R23}, Weyl
semimetals \cite{R24}, semiconductor quantum wells and
wires \cite{R25}, single trapped ion \cite{R26}, photonic
crystals \cite{R27}, ultracold atoms \cite{R28}, macroscopic sonic
crystals and photonic superlattices \cite{R29}, cavity
electrodynamics \cite{R30}, ultracold atoms in designed laser fields
\cite{R31,R31,R33,R34}, rolled-up electron gases \cite{R35},
electrons in quantum wells and dots in the presence of a magnetic
field \cite{R36}, spin-orbit-coupled Bose-Einstein
condensates \cite{R37}, photonic superlattices \cite{R38},
metamaterials \cite{R39}, Kitaev chain system \cite{R40}, black
holes \cite{R41}, photonic waveguide systems \cite{R42}, and mono and
bilayer graphene and carbon nanotubes \cite{R43}. This is because in
such structures two interacting bands give evidence of Dirac
equation for massless fermions in vacuum.\\
\indent This paper focuses on the elucidation of ZB effect in
relativistic quantum plasma. We assume the system is imbedded in a
neutralizing background of positive charge. To the best of our
knowledge, no deliberation has been made of the ZB effect in
quantum relativistic plasma using the combination of
Dirac wavefunctions and the Wigner operator. We here investigate the
trembling dynamics of Dirac's electrons based on the
Dirac-Heisenberg-Wigner(DHW) function \cite{R46}. This formalism
rests on the Wigner transform of the Dirac-Heisenberg correlation
function of the Dirac field operators. We exploit this approach for
the case of spin-1/2 Dirac particles in
relativistic quantum plasma in terms of phase-space concepts, which
are pretty easier and quite familiar, to
understand the nonclassical ZB effect. Since the discovery of
pulsars  in 1967 \cite{R47} and now owing to fabricating and
simulating Dirac materials \cite{R17}, the study of relativistic
quantum plasmas has been a highly desirable demand. With calculating the
expectation values of the position and velocity operators, we then
discuss the amplitude and frequency of the ZB oscillations for
Dirac electrons within quasi-one-dimensional relativistic quantum plasma.
We show that the oscillatory behavior of measurable quantities could be
associated with the Zitterbewegung effect which manifests itself
as a damping interference pattern stemming from mixing the
positive and negative dispersion modes of Dirac particles.\\
\indent%
The present paper is organized as follows. In next section, we
present our theoretical model for relativistic quantum plasma
based on Dirac's equation. In subsequent section, DHW
functions related to the field vacuum are utilized to assess the
ZB oscillations. Finally, in last section, we present the
conclusion.\\
\section{THEORETICAL MODEL}%
\label{sec2}%
\indent%
The Dirac's Hamiltonian for an electron with mass $m$ in the absence
of any external field is given by
\begin{equation}
\hat{H}=c\vec{\hat{\alpha}}\cdot\vec{\hat{p}}+\hat{\beta}mc^2
,\label{eq1}
\end{equation}
where $c$ is the speed of light, $\hat{\alpha}$ and $\hat{\beta}$
are the usual $4\times4$ Dirac matrices, and $\hat{p}$ is the
momentum operator. In Heisenberg picture one would make use of
$\vec{\hat{v}}(t)=d\vec{\hat{r}}/dt=(1/i\hbar)[\vec{\hat{r}}(t),\hat{H}]$
for the velocity $\vec{\hat{v}}$ and the position operator
$\vec{\hat{r}}=(x,y,z)$, respectively. Using Eq.(~\ref{eq1}) and by
applying the Baker-Hausdorff lemma we calculate the
commutator of the operators $\hat{H}$ and $\vec{\hat{r}}$ to obtain
the explicit result for the time evolution of the position operator
as below \cite{R48}
\begin{equation}
\vec{\hat{r}}_{H}(t)=\vec{\hat{r}}_{H}(0)+c^{2}{\hat{H}}^{-1}\vec{\hat{p}}t
-\frac{i}{2}\hbar c\vec{\hat{\alpha}}(0){\hat{H}}^{-1}\left[1-\exp\left(-\frac{i2\hat{H}t}{\hbar}\right)\right]
+\frac{i}{2}\hbar c^{2}{\hat{H}}^{-2}\vec{\hat{p}}\left[1-\exp\left(-\frac{i2\hat{H}t}{\hbar}\right)\right],
\label{eq2}
\end{equation}
where the first and second terms that are linear in time describe
the classical motion of the Dirac electron. The third and
fourth terms represent oscillatory movements which are of quantum
nature and could represent the ZB effect.\\
\indent To examine the relativistic effect like Zitterbewegung, we
calculate the expectation value of the position operator, i.e.,
$\langle\vec{\hat{r}}_{H}(t)\rangle$, for Dirac
electrons within the degenerate quantum plasma. In the context of
plasma physics, a statistical description of the systems could be
done using the usual approach of the Wigner formalism as follows
\begin{equation}
\langle\vec{\hat{r}}_{H}(t)\rangle=\int d\vec{r}\int d\vec{v}{\bf w}(\vec{r},\vec{v},t)\vec{\hat{r}}_{H}(t),
\label{eq3}
\end{equation}
where the relativistic Wigner distribution matrix ${\bf w}$ is generally a
$4\times4$ matrix constructed by $16$ components. Based on two sets,
$(1,\vec{\rho})$ and $(1,\vec{\sigma})$, with $2\times2$ Pauli
matrices including the unit matrix for completeness, the general
expansion of the DHW functions into the basis set of matrices forms
the Wigner matrix as follows \cite{R46}
\begin{equation}
{\bf w}(\vec{r},\vec{v},t)=\frac{1}{4}\left(f_{0}+\vec{\sigma}\cdot\vec{g}_{0}+\sum_{i=1}^{3}\rho_{i}(f_{i}+\vec{\sigma}\cdot\vec{g}_{i})\right),
\label{eq4}
\end{equation}
in which the coefficients $f$'s, with $f_0$ and $f_3$ as scalars and
$f_1$ and $f_2$ as pseudoscalares, and $\vec{g}$'s, with
$\vec{g}_{1}$ and $\vec{g}_{2}$ as vectors and $\vec{g}_{0}$ and
$\vec{g}_{3}$ as pseudovectors, are dimensionless functions of $\vec
r$, $\vec v$, and $t$. All DHW functions are real over the phase
space and some of them have clear physical meaning \cite{R46}. Here
we assume a constant chemical potential (Fermi energy) at zero
temperature, the Wigner function does not then depend on
space-time.\\
\indent%
In the absence of electromagnetic fields, with
$\hbar=c=k_{B}=1$, the only non-zero DHW functions for free Dirac
electrons are $f_{3}=-2m/E$ and $\vec{g}_1=-2\vec{p}/E$, where
$E=\sqrt{m^2+p^2}$. The Wigner matrix thus reads
\begin{equation}
w_{\alpha\beta} =\frac{1}{4}
 \begin{pmatrix}
  f_{3} & 0 & g_{1}^{z} & g_{1}^{-}\\
  0 & f_{3} & g_{1}^{+} & -g_{1}^{z}\\
  g_{1}^{z}  & g_{1}^{-}  & -f_{3} & 0\\
  g_{1}^{+} & -g_{1}^{z} & 0 & -f_{3}
 \end{pmatrix}
 ,
 \label{eq5}
\end{equation}
where ${g}_{1}^{z}=-2p_{z}/E$ and $g_{1}^{\pm}=p_{x}\pm ip_{y}$.
\section{RESULTS AND DISCUSSIONS}%
\label{sec4}%
\indent%
Electron correlation effects owing to Coulomb interaction can be strong
in quasi-one-dimensional nanostructured materials.
Also, electrons in different semiconductors experience environment with various dielectric
constant. For the sake of simplicity, we assume quasi-one-dimensional and elongated
quantum plasma with $p_{x}\ll m^{*}c$ and $p_{y}\ll m^{*}c$, which might be formed 
in half-wavelength silicon crystal\cite{R49, R50}. Here $m^{*}$ is the effective mass 
of the electrons in silicon\cite{R51}. We suppose also $p_{z}\in[-p_{F},p_{F}]$.
At zero-temperature, the 
quantum plasma is quite degenerate and the chemical potential is
equal to the Fermi energy $E_F$. Also, the Fermi momentum is given
by $p_{F}/m^{*}c=({3n}/{8\pi(m^{*}c/\hbar)})^{1/3}$, with $n$ to be the
electron number density of plasma. Therefore, the Fermi energy of
the plasma could be expressed by \cite{R51}
\begin{equation}
\frac{E_{F}}{m^{*}c^2}=\left[\left(\frac{3n}{8\pi}\right)^{\frac{2}{3}}\left(\frac{2\pi\hbar}{m^{*}c}\right)^{2}+1\right]^{\frac{1}{2}}-1.
\label{eq6}
\end{equation}
\begin{figure}[htb]
\includegraphics[scale=0.7]{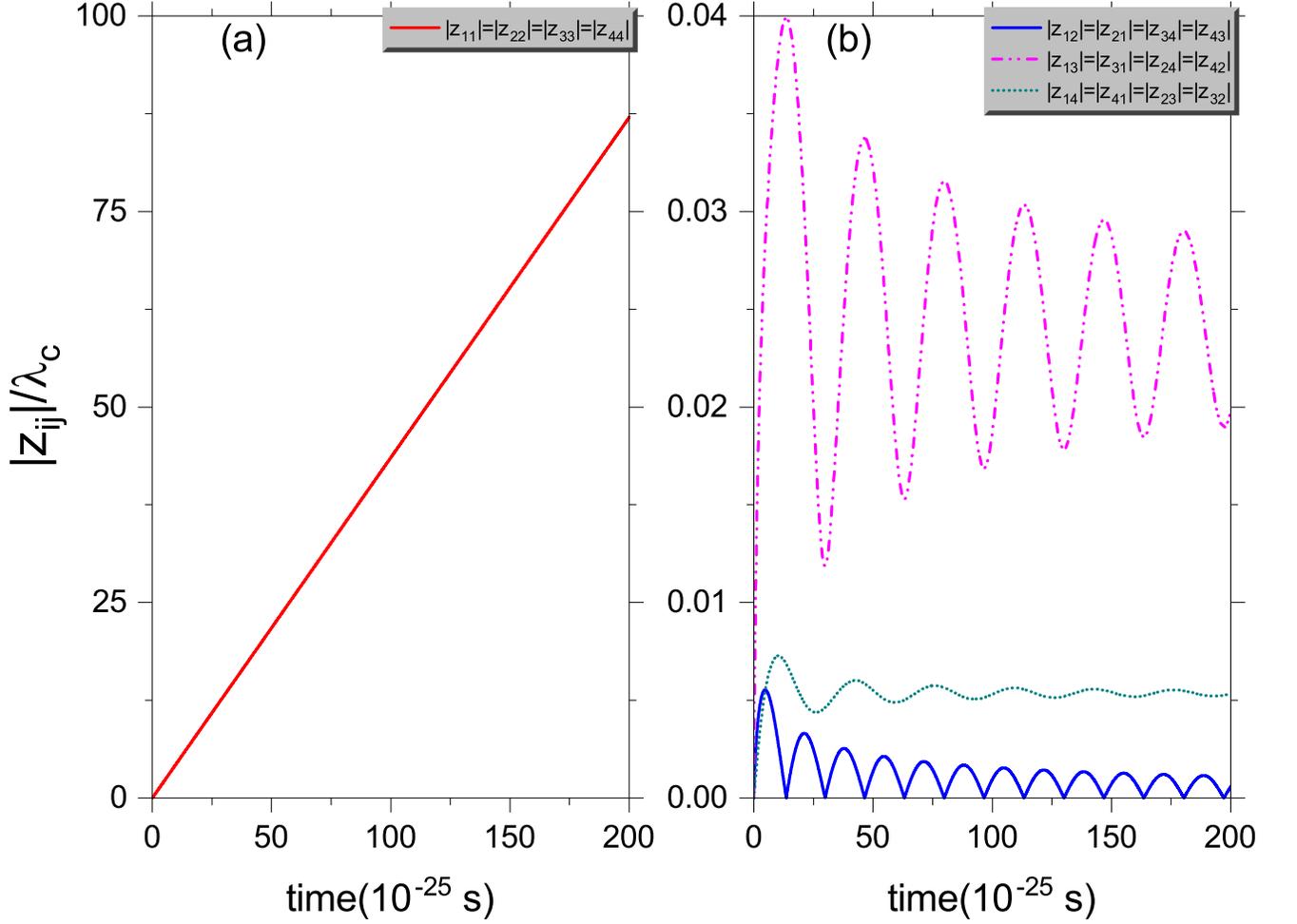}
{\caption{(Color online) ZB oscillations plots: Panels (a) and (b)
exhibit the normalized temporal oscillations of different components of
$\langle{{z}}_{H}^{\alpha\beta}(t)\rangle-$matrix given by
Eq.~(\ref{eq8}) for $|\langle{{z}}_{11}(t)\rangle|=|\langle{{z}}_{22}(t)\rangle|=|\langle{{z}}_{33}(t)\rangle|=|\langle{{z}}_{44}(t)\rangle|$(solid-red), with linear behavior, and
$|\langle{{z}}_{13}(t)\rangle|=|\langle{{z}}_{31}(t)\rangle|=|\langle{{z}}_{24}(t)\rangle|=|\langle{{z}}_{42}(t)\rangle|$(dashed-dot-dotted-magenta),
$|\langle{{z}}_{14}(t)\rangle|=|\langle{{z}}_{41}(t)\rangle|=|\langle{{z}}_{23}(t)\rangle|=|\langle{{z}}_{32}(t)\rangle|$(dotted-dark green),
$|\langle{{z}}_{12}(t)\rangle|=|\langle{{z}}_{21}(t)\rangle|=|\langle{{z}}_{34}(t)\rangle|=|\langle{{z}}_{43}(t)\rangle|$(solid-blue), with oscillating behavior, respectively. We choose the Compton
wavelength $\lambda_c$ as the normalization factor.} \label{Figure1}}
\end{figure}
\begin{figure}[htb]
\includegraphics[scale=0.6]{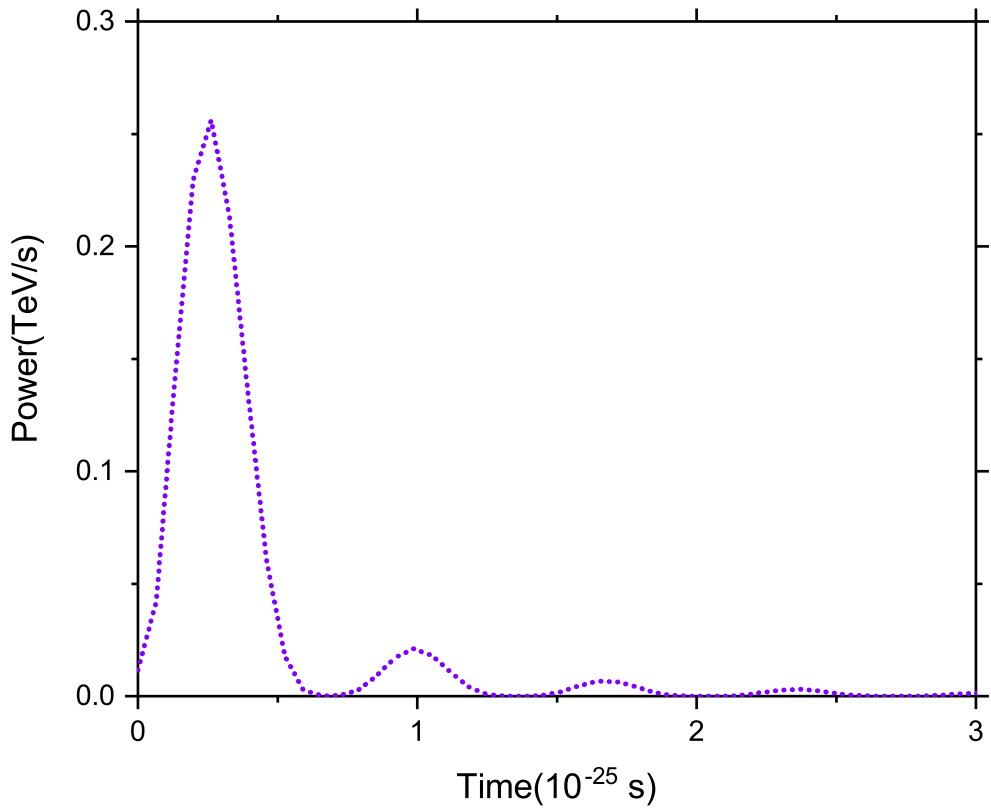}
{\caption{(Color online) The calculated total power radiated by Dirac electrons within
the plasma. As time passes, it rapidly decreases.}
\label{figure2}}
\end{figure}
\begin{figure}[htb]
\includegraphics[scale=0.6]{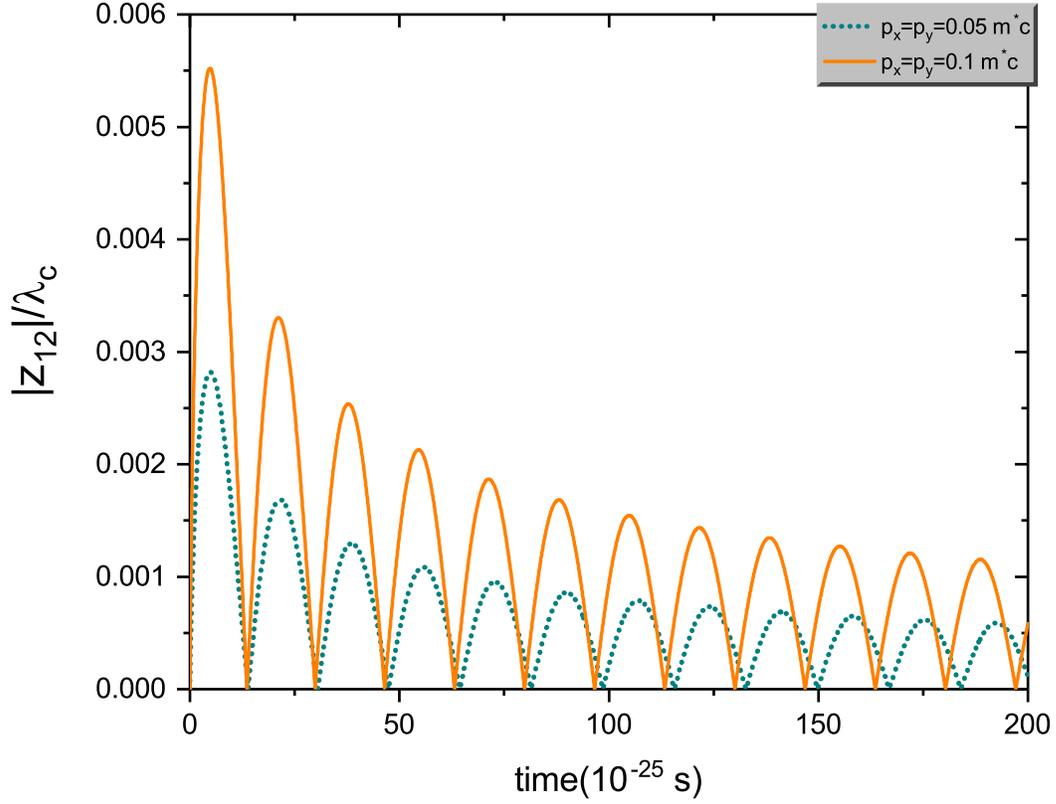}
{\caption{(Color online) The effect of transverse spatial extension of the plasma pipe
on the characteristics of ZB oscillations. The time evolution of the component
$z_{11}(t)$ for $p_{x}=p_{y}=0.05$ $m^{*}c$ (dark-cyan-dotted line) and
$p_{x}=p_{y}=0.1$ $m^{*}c$ (orange-solid line)is shown. With decreasing the radial spatial extension
of the one-dimensional plasma, the period of ZB oscillation diminishes.}
\label{figure3}}
\end{figure}
\begin{figure}[htb]
\includegraphics[scale=0.6]{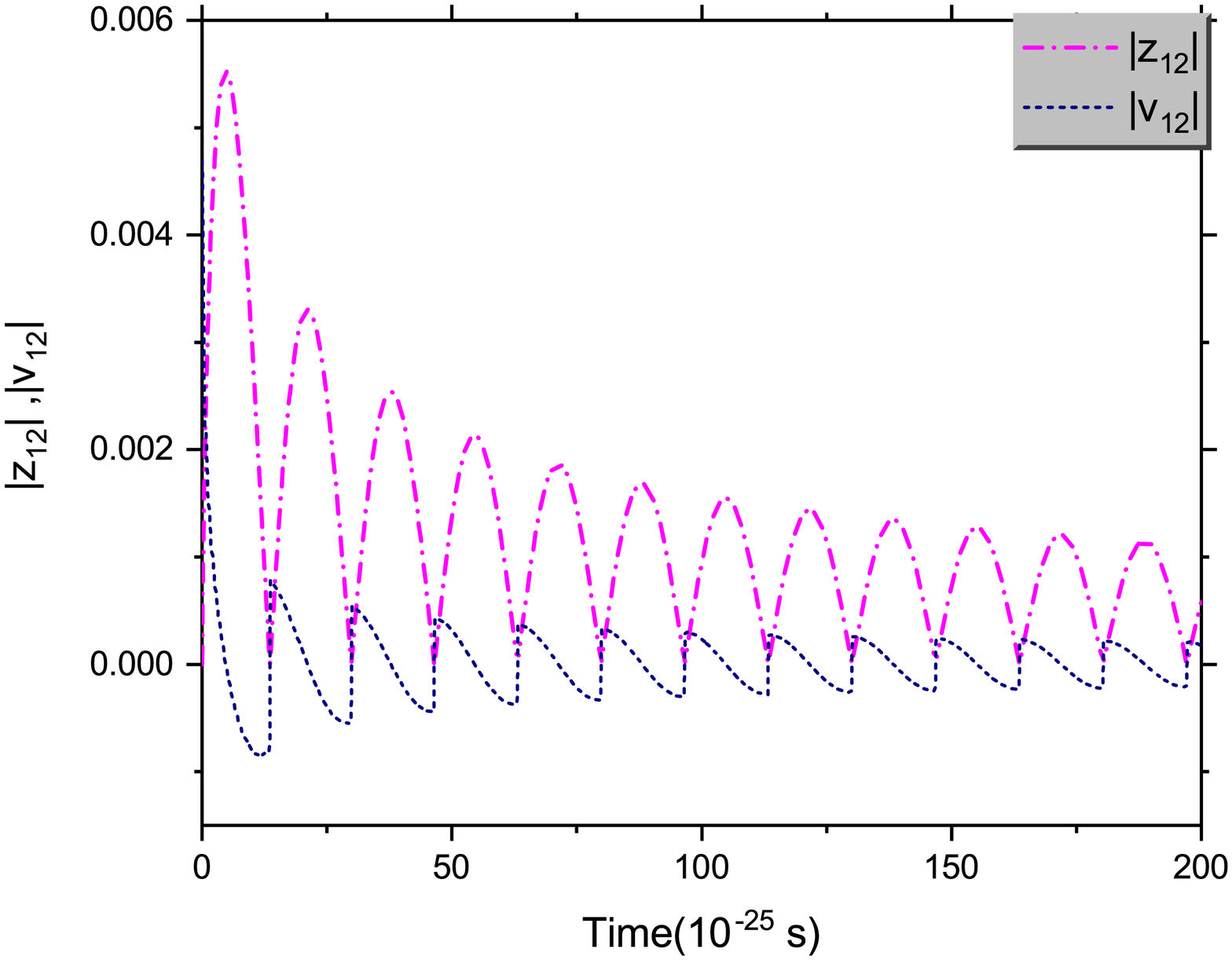}
{\caption{(Color online) The calculated $z_{11}(t)$ (magenta-dash-dotted line) and its corresponding velocity $v_{11}^{z}$ (navy-short-dashed line) as functions of time.}
\label{figure4}}
\end{figure}
In the relativistic quantum plasma the condition $E_{F}\gg m^{*}c^{2}$
is fulfilled. In the natural units
$\hbar=c=1$, since the Hamiltonian $\hat{H}$ is a $4\times4$ matrix,
the position operator $\vec{\hat{r}}_{H}(t)$ is a $4\times4$ matrix
as well. For example, upon doing some tedious algebra and via using the matrix form of Eq.(~\ref{eq2}) , one would
find the following expression in Heisenberg picture for the component $\hat{z}_{11}(t;p_x,p_y,p_z)$ as follows
\begin{equation}
\hat{z}_{H}^{11}(t;p_x,p_y,p_z)=\frac{p_{z}}{8}\frac{\left[(a^4-m^{*}a^3+ba^2)e^{-iat}-(a^4+m^{*}a^3-ba^2)e^{iat}+2m^{*}a^3\right]}{a^{5}\sqrt{a^{2}-m^{*2}}},
\label{eq7}
\end{equation}
where $a=\sqrt{p_{x}^{2}+p_{y}^{2}+p_{z}^{2}+m^{*2}}$ and $b=p_{x}^{2}+ip_{x}p_{y}+p_{z}^{2}+m^{*2}$.
To examine the ZB oscillations, by integration over $p_z$ at fixed values of
$p_x$ and $p_y$, we calculate the $4\times4$ matrix $\langle{{z}}_{H}^{\alpha\beta}(t)\rangle$, as a
complex function of time, given by
\begin{equation}
\langle{{z}}_{H}^{\alpha\beta}(t)\rangle=
 \begin{pmatrix}
  z_{11} & z_{12} & z_{13} & z_{14}\\
  z_{21} & z_{22} & z_{23} & z_{24}\\
  z_{31} & z_{32} & z_{33} & z_{34}\\
  z_{41} & z_{42} & z_{43} & z_{44}
 \end{pmatrix}
 ,
 \label{eq8}
\end{equation}
where every component $\langle z_{ij}\rangle$ is a complex function.\\
\indent
In the numeric calculations, for elongated one-dimensional quantum plasma, we choose the 
effective mass $m^{*}\simeq1.18~m$ \cite{R51}, with $m=0.5~MeV$. Also, we suppose the Fermi energy to be
$E_{F}\simeq 50$ $m^{*}c^{2}$. Therefore, the number density of plasma and its Fermi momentum 
become, $n\simeq1.2\times10^{44}/cm^{3}$ and $p_{F}\simeq 36$ $m^{*}c$, respectively. In the following we calculate
the amplitude of the components $|\langle{{z}}_{ij}(t)\rangle|=[(\Re{z_{ij}})^{2}+(\Im{z_{ij}})^{2}]^{1/2}$ given
by Eq.(~\ref{eq8}).\\
\indent%
In {Figure.1}, we exhibit the calculated ZB
oscillations of the components $|\langle{{z}}_{ij}(t)\rangle|$ as a
function of time. As shown, there are $16$ 
components which are classified by four groups
$|\langle{{z}}_{11}(t)\rangle|=|\langle{{z}}_{22}(t)\rangle|=|\langle{{z}}_{33}(t)\rangle|=|\langle{{z}}_{44}(t)\rangle|$(solid-red), with linear behavior, and
$|\langle{{z}}_{13}(t)\rangle|=|\langle{{z}}_{31}(t)\rangle|=|\langle{{z}}_{24}(t)\rangle|=|\langle{{z}}_{42}(t)\rangle|$(dashed-dot-dotted-magenta),
$|\langle{{z}}_{14}(t)\rangle|=|\langle{{z}}_{41}(t)\rangle|=|\langle{{z}}_{23}(t)\rangle|=|\langle{{z}}_{32}(t)\rangle|$(dotted-dark green),
$|\langle{{z}}_{12}(t)\rangle|=|\langle{{z}}_{21}(t)\rangle|=|\langle{{z}}_{34}(t)\rangle|=|\langle{{z}}_{43}(t)\rangle|$(solid-blue), with oscillating behavior
within panels $(a)$ and $(b)$, respectively. As shown in the panel $(a)$, the linear nature exhibits that the second term in Eq.(~\ref{eq2})
is dominant. On the other hand, in the panel $(b)$, the oscillatory style shows the ZB fluctuations.
The ZB oscillations are damped and their amplitudes
gradually decrease. One would associate this damping with the free
radiation. The moving free electron within plasma loses kinetic energy, that is
converted into photons. The power radiated by the moving electron in quasi-one-dimensional
plasma, in which the electron is either accelerated or decelerated
linearly, could be obtained by the relativistic and linear version
of Larmor's formula given by \cite{R53}
\begin{equation}
P(t)=\frac{e^{2}\gamma^{6}}{6\pi\epsilon_{0}c^{3}}a^{2}(t),
\label{eq9}
\end{equation}
where $\gamma$ is the usual Lorentz factor and $\vec{a}(t)=\vec{\ddot{r}}(t)$
stands for the acceleration of the mobile electron within plasma in the
$z-$direction. In Figure.~\ref{figure2}, the quantity of power $P(t)$ is sketched
as a function of time. It shows a sharp peak, which rises to a height
of $\sim0.25$ Tera electron Volt per second, and therefore swoftly diminishes.
In {Figure.1}(b), one can estimate the amplitude and period
of the ZB oscillation, for example, for the component $z_{11}(t)$ by
implementing Eq.(~\ref{eq7}). The frequency of ZB oscillations
is identified by $\omega_{ZB}=\sqrt{p_{x}^{2}+p_{y}^{2}+p_{z}^{2}+m^{*2}}$.
Owing to the great complexities involved in many-electron
systems as well as the limitations in experimental capabilities, it
is a challenge to manifest both the spatial extension and
the frequency of ZB oscillations for noninteracting and non-trapped
electrons \cite{R15}.\\
\indent%
Now we inspect the ZB oscillations with different widths of
the quasi-one-dimensional quantum plasma.
Figure.~\ref{figure3} presents the $|z_{12}(t)|$ for $p_{x}=p_{y}=0.05$
$m^{*}c$ (dark-cyan-dotted line) and $p_{x}=p_{y}=0.1$ $m^{*}c$ (orange-solid
line). With decreasing the transverse spatial extension of the
quasi-one-dimensional plasma cloud, the period of the ZB
oscillations, i.e., $T_{ZB}=1/\omega_{ZB}$, decreases.\\
\indent%
To acquire a measure of the velocity of electron within
 plasma, we compute it via $\vec{v}(t)=\vec{\dot{r}}(t)$. In
Figure.~\ref{figure4}, we chart the plot of $|z_{12}(t)|$ and its corresponding velocity
$|v_{12}^{z}(t)|\equiv|{\dot{z}}_{12}(t)|$ over a single panel. As shown, both the position
and velocity exhibit ZB oscillations, yet the amplitude of the velocity
curve approaches zero faster than the case of position. Moreover, the
oscillatory functioning of the velocity curve reveals that the ZB effect exposes
itself through a damping interference pattern coming from merging the
positive and negative modes of the Dirac particle dispersion relations.\\
\indent%
Finally,  it is worth noting that some elements of the Wigner matrix
expressed by Eq.(~\ref{eq5}) is negative which could be an
important token\emph{} of nonclassicality with application for
quantum information technologies \cite{R54,R55}.
\section{CONCLUSION}%
\label{sec5}%
\indent%
In this work, the relativistic single-particle Wigner function is
exploited to study the Zitterbewegung oscillations for
Dirac free particles in quasi-one-dimensional quantum plasma.
Although we inquire into in detail only a Dirac free particle propagating along the
$z-$direction, it is straightforward to infer that the ZB oscillations could occur
for all components of the position and velocity operators. This trembling
movement could be associated with the oscillatory behavior of the position operator which manifests itself as
an interference pattern originating from mixing the positive and negative
dispersion branches of the electron. The influence of electromagnetic
potentials as well as other trapping potentials in order to improve the amplitude and frequency of the ZB oscillations
will appear elsewhere.

\end{document}